\documentclass[aps,prl,twocolumn,groupedaddress]{revtex4}
\pdfoutput=1
\usepackage{graphicx}
\usepackage{color}
\usepackage{float}
\usepackage{subfigure}
\usepackage{dcolumn}
\usepackage{amsmath}
\usepackage{amssymb}

\def\be{\begin{equation}}
\def\ee{\end{equation}}
\def\bea{\begin{eqnarray}}
\def\eea{\end{eqnarray}}

\def\a{\alpha}
\def\l{\lambda}
\def\L{\Lambda}
\def\e{\epsilon}

\begin{document}

\title{Flat Dielectric Grating Reflectors with High Focusing Power}

\author{David Fattal}
\email{david.fattal@hp.com}
\affiliation{Hewlett-Packard Laboratories, 1501 Page Mill Road,
Palo Alto, CA 94304--1123}

\author{Jingjing Li}
\affiliation{Hewlett-Packard Laboratories, 1501 Page Mill Road,
Palo Alto, CA 94304--1123}

\author{Zhen Peng}
\affiliation{Hewlett-Packard Laboratories, 1501 Page Mill Road,
Palo Alto, CA 94304--1123}

\author{Marco Fiorentino}
\affiliation{Hewlett-Packard Laboratories, 1501 Page Mill Road,
Palo Alto, CA 94304--1123}

\author{Raymond G.\ Beausoleil}
\affiliation{Hewlett-Packard Laboratories, 1501 Page Mill Road,
Palo Alto, CA 94304--1123}


\maketitle

\textbf{Sub-wavelength dielectric gratings (SWG) have emerged recently as a promising alternative to distributed-Bragg-reflection (DBR) dielectric stacks for broadband, high-reflectivity filtering applications. A SWG structure composed of a single dielectric layer with the appropriate patterning can sometimes perform as well as thirty or forty dielectric DBR layers, while providing new functionalities such as polarization control and near-field amplification. In this paper, we introduce a remarkable property of grating mirrors that cannot be realized by their DBR counterpart: we show that a non-periodic patterning of the grating surface can give full control over the phase front of reflected light while maintaining a high reflectivity. This new feature of dielectric gratings could have a substantial impact on a number of applications that depend on low-cost, compact optical components, from laser cavities to CD/DVD read/write heads.}

Resonant effects in dielectric gratings were first clearly identified in the early 1990's~\cite{ref:wang1990gmr} as having promising applications to free-space optical filtering~\cite{ref:wang1993tag,ref:magnusson1992npo} and sensing~\cite{ref:magnusson2005crl,ref:fattal2007gmr}. They typically occur in sub-wavelength gratings (SWG), where the first-order diffracted mode corresponds not to freely propagating light but to a guided wave trapped in some dielectric layer. The trapped wave is scattered into the zeroth diffracted order and interferes with the incident light to create a pronounced modulation of transmission and reflection. When a high-index-contrast grating is used, the guided waves are rapidly scattered and do not propagate very far laterally. In this case it is appropriate to think of the grating as a coupled resonator system, where each high index groove behaves as a (lossy) cavity. In such gratings, broad transmission and reflection features can be observed, and have been used to design novel types of highly reflective mirrors~\cite{ref:mateus2004umu,ref:mateus2004bbm}. Recently, SWG mirrors have been used to replace the top dielectric stacks in vertical-cavity surface-emitting lasers (VCSELs)~\cite{ref:huang2007sel,ref:huang2008smh}, and in novel MEMS devices~\cite{ref:jung2007hrb,ref:huang2008ntl}. Besides being more compact and cheaper to fabricate, these structures provide new optical features such as polarization control.


In this paper, we show --- both numerically and experimentally --- how a non-periodic design of the SWG pattern allows dramatic control of the phase front of the reflected beam, \emph{without} affecting the high reflectivity of the mirror. We provide complete design rules on how to obtain a given phase front for the reflected beam, and we design a set of effectively cylindrical and spherical mirrors which can be fabricated using simple lithography followed by etching. We believe our method can considerably reduce the cost of optical components while offering new functionalities for a host of applications from laser cavity reflectors to CD/DVD read/write heads. The paper will focus on 1D gratings with TM-polarized incident light, but the design rules apply equally well to TE-polarized and unpolarized gratings (the latter necessitating a 2D pattern). Transmissive gratings are also possible and will be described elsewhere.

\begin{figure}
    \centering
    \subfigure[]{
    \includegraphics[width=3in]{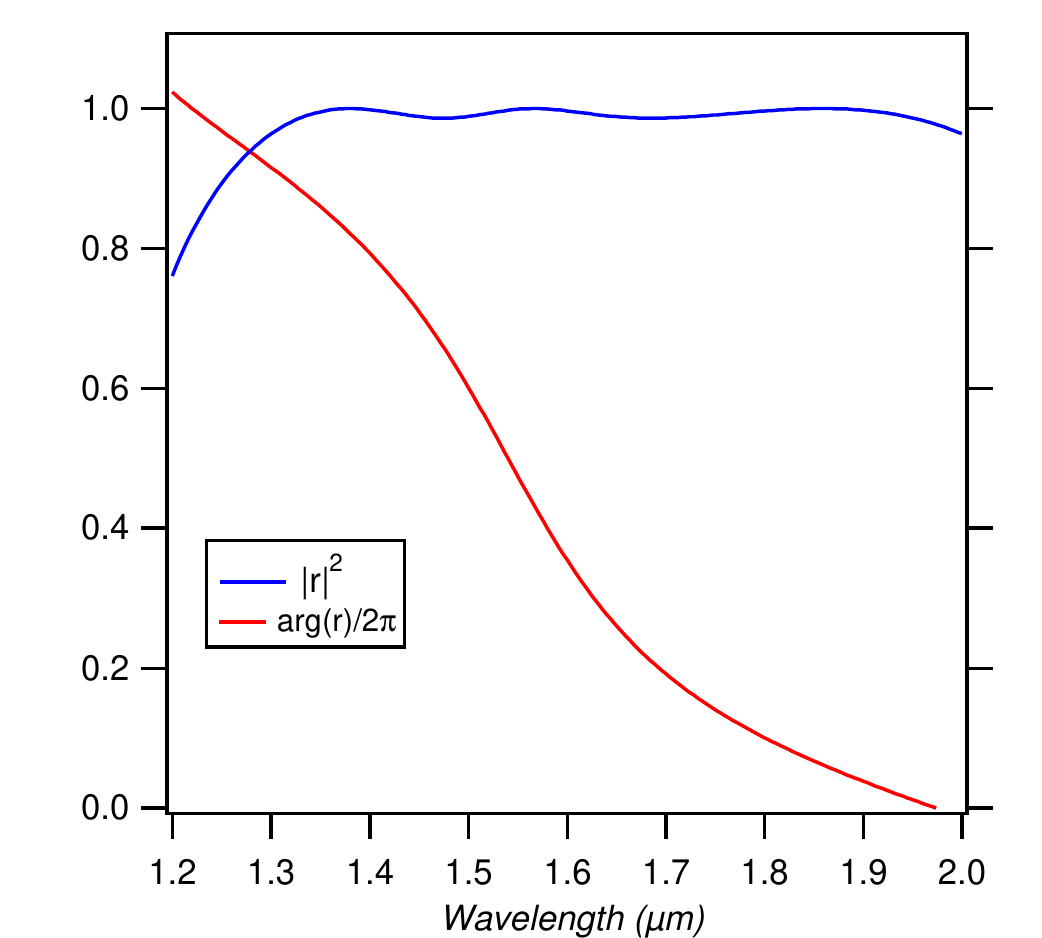}
    \label{fig:refl_vs_lam}}
    \hspace{.1in}
    \subfigure[]{
    \includegraphics[width=3in]{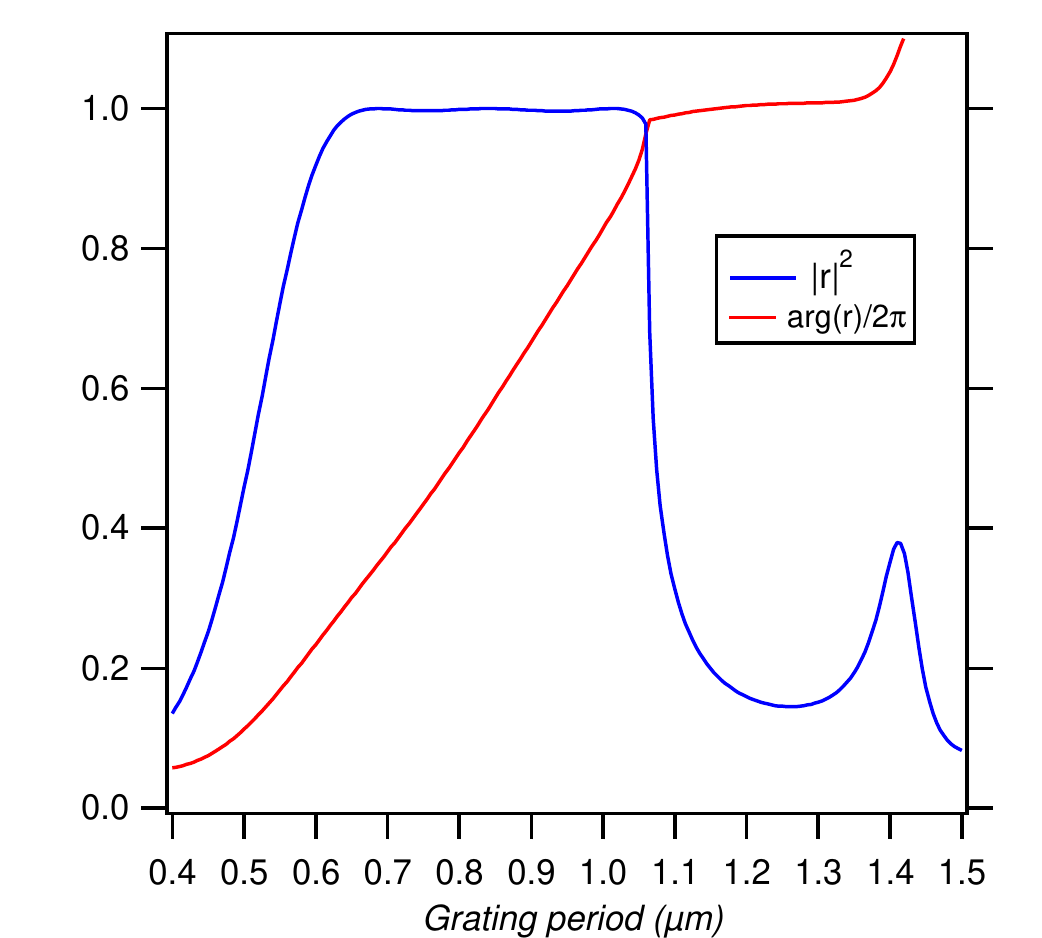}
    \label{fig:refl_vs_a}}
    \caption{Magnitude-squared and phase of the reflection coefficient from a 1D silicon grating of 470~nm thickness on a $\textrm{SiO}_2$ substrate, subject to normal incidence illumination with a TM-polarized plane wave. The reflection coefficient is plotted as (a) a function of wavelength for a fixed period of 0.76~$\mu$m, and (b) as a function of spatial period at a fixed wavelength of 1.55~$\mu$m.}
\end{figure}

To understand the phase front modification provided by a \emph{non-periodic} SWG mirror, we start by considering the complex reflection coefficient $r_0(\lambda)$ of a particular \emph{periodic} SWG mirror, as shown in Fig.~\ref{fig:refl_vs_lam}. The wavelength-dependent reflectance and phase shift of the mirror can be easily obtained using either the finite element method (FEM) or rigorous coupled wave analysis (RCWA)~\cite{ref:li1997nff}. In the particular case of Fig.~\ref{fig:refl_vs_lam}, the grating is a set of linear silicon grooves with a period of 0.76~$\mu$m on a quartz substrate, and is illuminated at normal incidence with a TM-polarized plane wave (i.e., the $E$-field vector is perpendicular to the grooves). Due to the strong index contrast between silicon and air, the grating has a broad spectral region of high reflectivity, consistent with other reports in the literature\cite{ref:mateus2004umu}. A less obvious but critical observation is that the phase of the reflected beam varies significantly across the high-reflectivity spectral region.

In order to design grating reflectors that imprint a particular phase profile on an incident beam, we first exploit a well-known property of Maxwell's equations with respect to a uniform spatial scale transformation. If we physically change the spatial dimensions of a periodic grating \emph{uniformly} by a factor of $\alpha$, the new reflection coefficient profile will be identical to that of the original grating, but with a wavelength axis that has also been scaled by $\alpha$. Therefore, if we design a grating that has a particular complex reflection coefficient $r_0$ at a vacuum wavelength $\lambda_0$, then we obtain a new grating with the same reflection coefficient at wavelength $\lambda$ by multiplying all grating dimensions by the factor $\alpha = \lambda/\lambda_0$, giving $r(\lambda) = r_0(\lambda/\alpha) = r_0(\lambda_0)$.

The critical conceptual advance that enables designs with high focusing power is the realization that a \emph{non-uniform} scaling of the reference grating should allow local tuning of the value of the reflected phase while maintaining a large reflectivity for a monochromatic incident beam. Since high-contrast gratings operate with \emph{localized} resonances, we expect their reflection properties at a given point in space to depend only on the local geometry around that point. Suppose that we wish to imprint a phase $\phi(x, y)$ on a reflected beam at some point on the grating with transverse coordinates $(x, y)$. Near that point, a nonuniform grating with a slowly-varying spatial scale factor $\alpha(x, y)$ behaves locally as though it was a periodic grating with a reflection coefficient $r_0[\lambda/\alpha(x,y)]$. Therefore, given a periodic grating design with a phase $\phi_0$ at some wavelength $\lambda_0$, choosing a local scale factor $\alpha(x,y) = \lambda/\lambda_0$ will set $\phi(x, y) = \phi_0$. So long as all required values of $\phi_0$ are available within the high-reflectivity spectral window, the new non-periodic grating can be designed to support a specific phase map for a specific application. We have followed this procedure to generate the plot of the magnitude and phase of the complex reflection coefficient at $\lambda = 1.55~\mu$m as a function of grating period shown in Fig.~\ref{fig:refl_vs_a}. Notice that this approach is in general not valid for weak gratings where the local structure of the incident beam is spatially averaged laterally upon reflection~\cite{ref:avrutsky1989rbf}. However, as we show below, this is a very robust and reliable guiding principle for high contrast grating designs.

Two simple phase profiles are noteworthy. First, a linearly varying phase profile $\phi(x,y)=\kappa_x x + \kappa_y y$ (where $\kappa_x$ and $\kappa_y$ are design parameters) gives rise to a deflected reflected beam, just as a tilted mirror would.  The deflection angle $\theta$ measured from normal is given by $\sin\theta= \sqrt{\kappa_x^2+\kappa_y^2}/k_0$, where $k_0=2\pi/\lambda_0$ and $\lambda_0$ is the free space wavelength. Second, a quadratic phase profile $\phi(x,y)= k_0\left( x^2/f_x + y^2/f_y \right)/2$ gives rise to elliptical ``thin lens'' focusing with focal lengths $f_{x}$ and $f_{y}$ in the $xz$ and $yz$ planes, respectively. In that case, the grating behaves as a perfect parabolic mirror. Of course, more complicated reflection phase profiles can also be implemented.

Continuously scaling the height of a particular grating groove typically requires grey-scale lithography. While this is a mature fabrication technique, it is not readily accessible to most research groups and is more expensive than binary lithography. For this reason, whenever possible we seek purely planar designs for our SWG reflectors, maintaining the grating groove thickness across the entire optic. In practice, it is generally possible to obtain a phase shift similar to that achievable in 3D scaling using purely 2D scaling, varying the grating period and/or duty cycle (defined as the groove width divided by the period) only. Figure~\ref{fig:refl_vs_a} shows an example where a phase shift close to $2\pi$ is obtained for a fixed wavelength by varying the grating period only while keeping the reflectance above $98\%$. The resulting design is reminiscent of output-coupler gratings used to extract light from a dielectric slab into a controlled radiation mode~\cite{ref:nishiwaki1994ohe,ref:eriksson1996hdg,ref:eriksson1996ebd}. However, the operating principle is very different here: it is based on the excitation of localized resonances in the high-contrast grating rather than on leaky modes in a low-contrast grating. It is also very different from (and superior to) conventional Fresnel lenses, where prisms and wedges of well-engineered geometric features are utilized to exploit refraction and total internal reflection. In contrast to Fresnel lenses, SWGs have only sub-wavelength features that provide a smooth reflected phase front despite the sharp discontinuities in dielectric constant.

Once we have chosen a particular spatial distribution of duty cycle and period, we must determine the grating pattern that best approximates that distribution. One way to think about this problem is to view the non-periodic grating as a distorted version of the reference periodic grating. A local change in duty cycle simply corresponds to a local scaling of the grating pattern that leaves the underlying lattice unchanged, a procedure that is easily implemented for 1D or 2D gratings. A local change of period can be implemented by a coordinate transformation that distorts the lattice itself. A more detailed discussion of this idea can be found in the Appendix section.

Our freedom in designing optical devices using purely planar technology is determined by the range of phase variation we can obtain through lateral scaling of a dielectric grating.  Many applications --- including phase front correction, laser cavity reflectors, and low-numerical-aperture optics --- require only a relatively small phase variation (e.g., less than $\pi/2$) achievable through the lateral scaling of high-contrast gratings having a \emph{single resonance}. For example, in the case of the planar ``parabolic'' reflectors described below, a phase variation of $0.8\pi$ at a fixed operating wavelength can be realized by varying the duty cycle alone. Larger phase differentials can be obtained using dielectric gratings that exhibit \emph{merged multiple resonances} in their spectrum, such as the 470~nm-thick silicon grating on an SiO$_2$ substrate operating in the 1.2--2~$\mu$m wavelength range shown in Fig.~\ref{fig:refl_vs_lam}. Applying the design principle outlined above, Fig.~\ref{fig:refl_vs_a} demonstrates that a total phase range of $1.7\pi$ can be realized at an operating wavelength of 1.55~$\mu$m by varying both duty cycle and local grating period, while the reflectance remains above 98\%. Although this phase variation is already large enough to support a wide range of free-space optical applications, achieving a full $2\pi$ is important because it enables arbitrary phase front control. A straightforward means of reaching this significant goal is to use a few (as little as two) grating thicknesses with complementary phase variations that together cover the entire $2\pi$ range. This allows a reflectance in excess of 98\% to be maintained using a process compatible with planar technology.  Another solution (used next to simulate a large-NA parabolic reflector) arises from the observation that we can realize a $2\pi$ phase variation (and thus full wavefront control) using a single-thickness design by accepting a small decrease in the reflectance.

\begin{figure*}
    \includegraphics[width=5in]{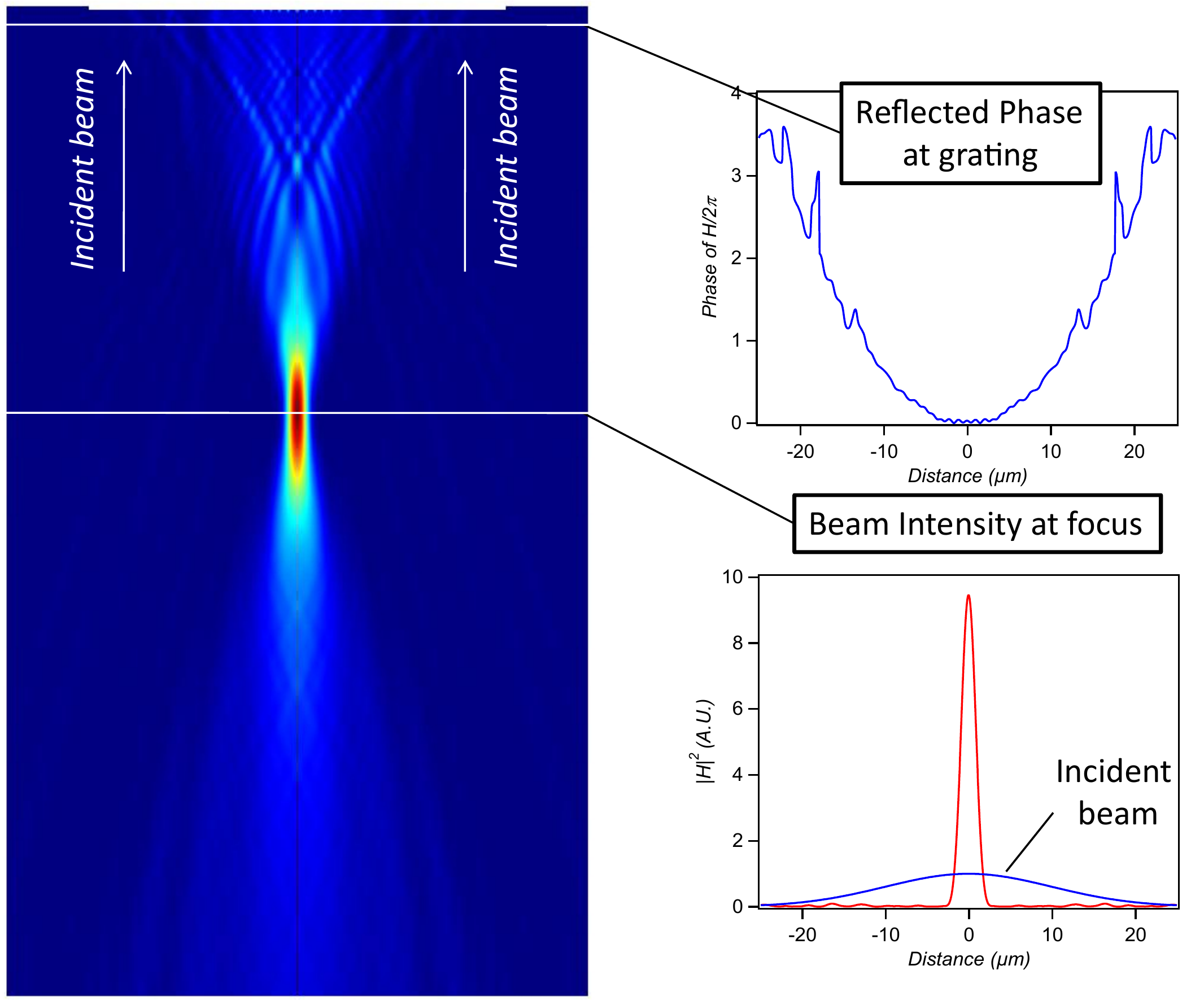}
    \caption{Numerical simulation of a 50~$\mu$m-aperture focusing reflector with NA = 0.45, featuring an $8\pi$ differential phase shift from center to edge. The grating is illuminated with a collimated beam of waist 20~$\mu$m at a wavelength of 1.55~$\mu$m. The plot shows the intensity of the reflected field as a function of position, as well as both the phase distribution 1~$\mu$m below the grating and the beam intensity profile in the focal plane. The beam is almost perfectly gaussian at the focus with a waist of 1.66~$\mu$m.  }
    \label{fig:large_NA}
\end{figure*}

Figure~\ref{fig:large_NA} provides an example of arbitrary wavefront control with reasonably high reflectance using lateral scaling alone. Here we have used a finite element method to simulate a 1D (cylindrical) parabolic reflector of large numerical aperture ($\textrm{NA} = 0.45$) with a diameter of 50~$\mu$m and a focal length of 50~$\mu$m .  The design of the device requires a total phase variation of $8 \pi$ from center to edge, which is obtained by modulation of the local period only using the reflectance curve of Fig.~\ref{fig:refl_vs_a}. As shown in Fig.~\ref{fig:large_NA}, this optic focuses an incident gaussian beam with a 20~$\mu$m waist at the grating surface down to a spot with a 1.66~$\mu$m waist (i.e., within 25\% of the diffraction limit), with a total reflectance of 85\%. This rational design method constitutes an excellent starting point for a systematic optimization of the groove width and period, which is very likely to improve the amplitude and beam quality of the reflected field.

\begin{figure}
    \includegraphics[width=3.5in]{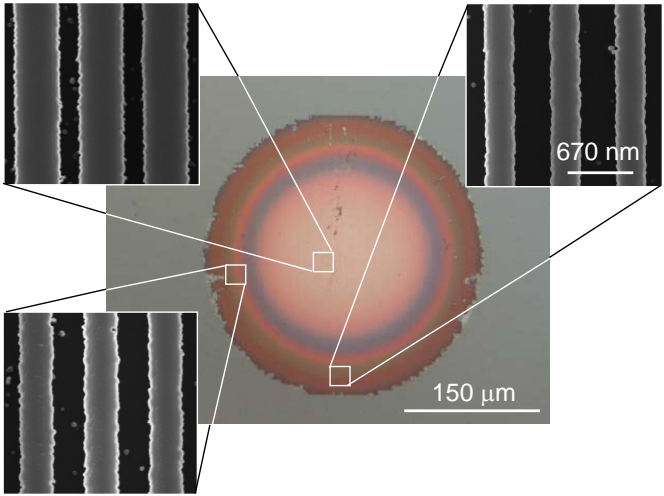}
    \caption{Optical microscope picture of a fabricated spherical SWG mirror. The groove width in various locations is shown as SEM images in the insets.}
    \label{fig:pics}
\end{figure}

\begin{figure*}[htbp]
    \includegraphics[width=6in]{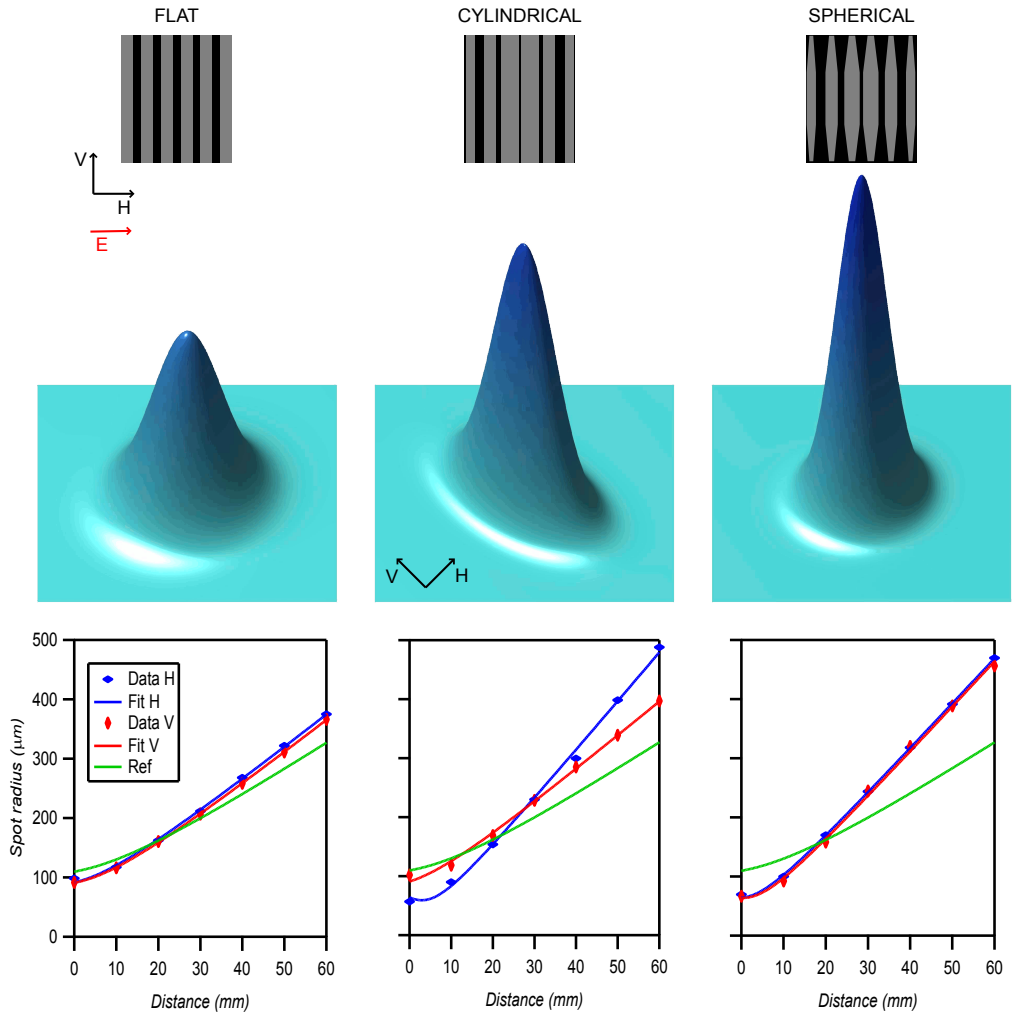}
    \caption{Top row: schematic of the groove distribution for the three structures: flat (left), cylindrical (middle), and spherical (right) SWG mirrors. Middle row: measured beam profiles at the foci for the three mirrors.  Bottom row: plots of the beam radii ($1/e^2$) as a function of distance for the three mirrors.  The blue and red diamonds are the measured values for the horizontal and vertical radii, respectively, and the continuous lines are best fits.  The green line represents the radius of a beam reflected from a reference plane mirror.}
    \label{fig:results}
\end{figure*}

We have tested the non-periodic SWG concept experimentally by fabricating TM reflectors with aperture radii of 150~$\mu$m (limited by our $e$-beam lithography tool) and a relatively long focal length of 17.2~mm to facilitate the measurement of the beam profile. This design required a phase variation of up to $\sim0.8\pi$ that was obtained through spatial modulation of the duty cycle alone, as described in detail in the Appendix section.  We fabricated a flat, a cylindrical, and a spherical SWG mirror made of a 450~nm amorphous silicon layer on a quartz substrate with a (uniform) grating period of 670~nm.  The fabrication procedure involving $e$-beam lithography is described below in the Methods section. The groove width is uniform for the flat reflector but is spatially modulated for the cylindrical and spherical mirrors. An optical microscope image of a fabricated spherical grating mirror is shown in Fig.~\ref{fig:pics}, along with scanning microscope images of the silicon grooves at various locations. We tested the device with a collimated laser beam having a 100~$\mu$m waist radius. Our main experimental results are summarized in Fig.~\ref{fig:results}, with the schematics of the grating patterns shown in the top row, and the measured beam profiles after reflection off each grating displayed in the middle row. The reflected beam parameters can be reconstructed from the measured beam radii at various positions, as shown in the bottom row of Fig.~\ref{fig:results}. Using these parameters, we can calculate the focal lengths to be $20 \pm 3$~mm for both mirrors, close to the design value of 17.2~mm.  The reflectance of the mirrors is in the range of 80--90\%, lower than the expected 98\%, primarily due to proximity effects in the $e$-beam lithography step, as well as the surface roughness of the silicon grooves evident in Fig.~\ref{fig:pics}. We believe that focal length and reflectance will be much closer to the design specifications after optimization of the fabrication procedure.

The non-periodic SWG mirrors presented here could provide new optical functionalities for a host of passive or active devices. They can provide tight lateral confinement of light in a VCSEL cavity, improving on the periodic-SWG-based structure demonstrated in \cite{ref:huang2008smh} and allowing high-speed, high-power, single-mode operation.  They can also be used to shape directly the transverse mode profile of a microlaser through the output mirror, replace the expensive compound lens systems in a number of consumer electronic products (DVD players, digital cameras, etc), and provide a cheaper alternative to micro-lens arrays used in CMOS and CCD sensors and quantum computation implementations relying on trapped atoms.  Combined with techniques to modulate the position, shape, or effective index of the grating layer (electro-optic effect, MEMS tuning, free carrier plasma effect, etc.), the approach presented here could become the tool of choice for applications involving active phase front correction, dynamic focusing, or cavity resonance tuning.

\section{Methods}

The devices of Figures \ref{fig:pics} and \ref{fig:results} were fabricated in 450~nm thick amorphous silicon deposited on a quartz substrate at $300\, ^{\circ}$C by PECVD. The refractive index of the silicon layer was measured by ellipsometry over the wavelength range of interest, and those measurement values were used in the numerical simulations. The grating patterns were defined by electron beam lithography using the negative resist FOX-12 from Corning exposed at 200~$\mu$C/cm$^2$ and developed for 3 minutes in a solution of MIF~300. After development, the patterns were weakly descummed using CF4/H2 reactive ion etching to clear the resist residue between the grating lines. The silicon grooves themselves were formed by dry etching using a HBr/O2 chemistry. At the end of the process, a 100~nm resist layer remains on top of the silicon grooves and was included in the numerical simulations. Scanning electron microscope images of the fabricated grating are shown in Fig.~\ref{fig:pics}.

The measurement setup consisted of a fiber-coupled laser that was collimated using an aspheric lens to have a waist of 100~$\mu$m at the grating position.  The reflected beam was picked off using a cube beam-splitter.  To obtain the reference data, a planar dielectric mirror with reflectivity $>$99\% was used instead of the grating.  Reflectance data were also confirmed using a spectroscopic ellipsometer.

\section{Appendix}

This appendix section provides detailed discussions of a few technical points raised in the main text. First, we explain how to design a grating pattern to imprint a specific phase profile on a reflected beam, in both 1D and 2D. Second, we present more materials on the design of the fabricated reflectors of Fig.~3 and 4.

\subsection{Designing a non-periodic grating pattern for a specific reflected phase profile}

In this section, we explain the procedure for designing the pattern of a high-index-contrast non-periodic grating to imprint a particular phase profile $\phi(x,y)$ on a reflected beam.  For a given periodic grating, the magnitude and phase of the reflection coefficient $r$ at a given wavelength $\l$ will vary with the period $p$ and the duty cycle $\eta$ (defined as the ratio of the groove width to the period).  For certain grating thicknesses, we can achieve a high reflectivity (i.e., $|r|^2 > 98\%$) with a wide variation of phase over a relatively large range of values of $\{p, \eta\}$, as indicated by the closed solid line in Fig.~\ref{fig:phasecontour}.  In general, this result serves as a look-up table to determine the distribution of duty cycle period $p(x,y)$ and $\eta(x,y)$ according to the target phase distribution $\phi(x,y)$. Here we provide a detailed explanation of how to determine $p(x,y)$ and $\eta(x,y)$ in practice.

\subsubsection{Designing a 1D non-periodic grating pattern using a discrete algorithm}

\begin{figure*}
    \includegraphics[width=6in]{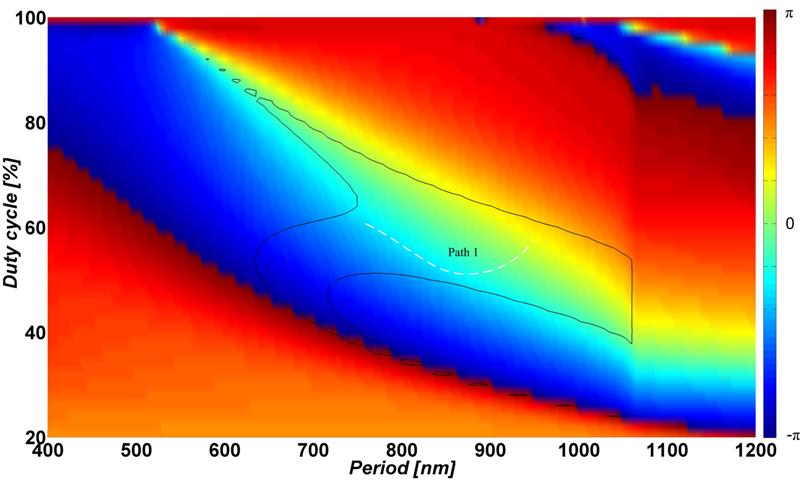}
    \caption{A plot of the phase of the reflection coefficient $r$ of a periodic SWG mirror as a function of period and duty cycle, used to design the large numerical aperture reflector shown in Fig.~2 in the main text.  The area enclosed within the solid line corresponds to $|r|^2>98\%$.}
    \label{fig:phasecontour}
\end{figure*}

For a 1D non-periodic grating design, the goal is to find the local period $p(x)$ and duty cycle $\eta(x)$ that give a particular known target phase profile $\phi(x)$ along the $x$ axis.  We begin by choosing a path (such as Path $1$ in Fig.~\ref{fig:phasecontour}) enclosed within the high reflectivity region, defined by a set of parametric functions of the parameter $t$ as
\begin{equation}
\label{eqn:LineEquation}
p = P\left( t \right), \quad \eta = I\left( t \right), \quad \phi = \Phi \left( t \right),
\end{equation}
where $P(t)$ and $I(t)$ are known, single-valued functions that generate coordinates on the horizontal and vertical axes, respectively, of Fig.~\ref{fig:phasecontour}, and $\Phi(t)$ generates the phase of the reflection coefficient (encoded by color in the figure). Given our target phase at the origin, $\phi(0)$, we select the value of $t \equiv t_0$ that gives $\Phi(t_0) = \phi(0)$, and record the corresponding period $p_0 = P(t_0)$ and duty cycle $\eta_0 = I(t_0)$.

Our task now is to use Fig.~\ref{fig:phasecontour} to create a discrete representation of the target function $\phi(x)$ by choosing the position and width of consecutive grating lines using the a straightforward recurrence procedure. (As noted in the main text, the physics of high-index-contrast subwavelength gratings results in a smoothed spatial distribution of phase with local values interpolated between adjacent grooves.) Suppose that we have already found the period and duty cycle of grating line $n$ to be $\eta_n \equiv I(t_n)$ and $p_n \equiv P(t_n)$, respectively, for some value of the parameter $t = t_n$, and that the center of that line is at $x_n$.  The center position of line $n + 1$ (with period $p_{n+1}$) can then be written as
\begin{equation}
\label{eqn:xnIteration}
{x_{n + 1}} = {x_n} + \frac{1}{2}\left(p_n + p_{n + 1}\right) .
\end{equation}
Therefore, we must find the value of $t_{n+1}$ that satisfies $\Phi(t_{n+1}) = \phi(x_{n+1})$, or
\begin{equation}
\label{eqn:tnIteration}
\Phi(t_{n+1}) = \phi \left[ x_n + \frac{1}{2} p_n + \frac{1}{2} P(t_{n+1}) \right] ,
\end{equation}
which depends on only one unknown $t_{n+1}$, and can be iterated numerically. The period and duty cycle of grating line $n+1$ are then computed as $p_{n+1}=P(t_{n+1})$ and $\eta_{n+1}=I(t_{n+1})$, and this process can be repeated until the boundary of the grating aperture is reached.

\subsubsection{Local period modulation of a 1D non-periodic grating through a coordinate transform}


As we described in the main text, the local period modulation of a 1D non-periodic grating can be realized analytically as a coordinate transformation from the reference periodic grating described by the dielectric permittivity distribution $\epsilon(u)=\epsilon(u+\Lambda)$. Consider a target period distribution $p(x)$, and define $\alpha(x) \equiv p(x)/p_0$, where $p_0$ is the period of the reference grating.  For simplicity, we assume that $\a(x)$ is piecewise continuous and is bounded, i.e. $0<A\leq \a(x) \leq B < +\infty$, which is always the case in practice.  We then define the coordinate transformation
\be u(x) = u(0) + \int_0^{x} \frac{ds}{\a(s)} \ee
By our assumptions on $\a(x)$, this map is continuously differentiable everywhere in $x$-space with
\be u^{\prime}(x) = \frac{1}{\a(x)} \ee
Consider a grating in $x$-space defined by the dielectric function $\tilde{\e}(x) = \e(u(x))$. When $\a(x)$ varies slowly compared to $\L$, we can write
\be u(x+\a(x)\L) \approx u(x) + \a(x)\L u^{\prime}(x) = u(x) + \L , \ee
which leads to the relation
\be \tilde{\e}(x+\a(x)\L) \approx \tilde{\e}(x) . \ee
That is, the grating defined in $x$-space by the dielectric function $\tilde{\e}$ has a local period $\a(x)\L$, so that $\tilde{\epsilon}(x)$ is indeed the grating pattern we are seeking.

\subsubsection{Designing a 2D non-periodic grating pattern}

As in the 1D case, the 2D grating pattern should be designed by using the magnitude/phase plot of the periodic grating design in Fig.~\ref{fig:phasecontour} as a look-up table.  For example, the discrete procedure for the 1D case can be repeated for discrete $y_n$ to determine the position of each period center $(x_n, y_n)$ with its duty cycle $\eta_{n,n}$.  The corresponding grid points of neighboring $y_n$ are then connected to form the non-uniform grating grooves with the local width determined by $\eta_{n,n}$.  Since the period at $y_n$ and $y_{n+1}$ is in general not the same, it is possible that the grids at $y_n$ and $y_{n+1}$ are not one-to-one matched, resulting in terminated grooves at a few locations similar to the dislocation defects in a crystal.  The density of such defects is low under slow-varying conditions, and their influence on the reflected field is localized, similar to the situation in the large NA reflector (shown in Fig.~2 of the main text) where discontinuities at the $2\, m\, \pi$ interfaces cause no noticeable effects in the final result.  The possible negative consequences of these defects can be further compensated by local modulation of the duty cycle.

%
\begin{figure*}
    \includegraphics[width=6in]{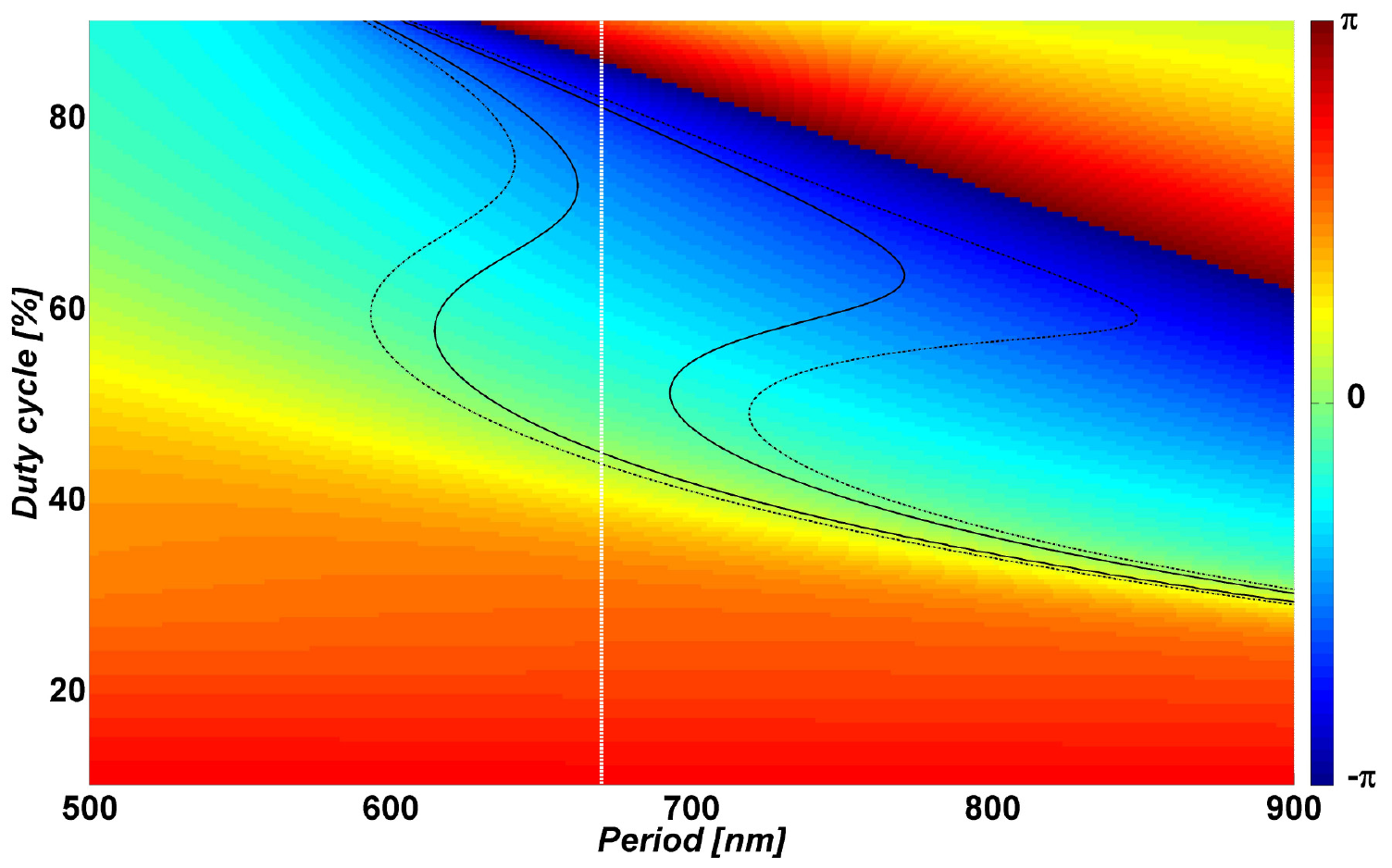}
    \caption{Simulated reflected phase of a periodic SWG mirror as a function of period and duty cycle, used to design the fabricated structures of Fig.~3 and 4 in the main paper. The grating is fabricated in $450$~nm thick silicon on a fused silica substrate, and operates at a wavelength of 1550~nm. The dashed and solid lines mark the contours for $|r|^2=95\%$ and $98\%$, respectively.}
    \label{fig:map_refl}
\end{figure*}

\subsection{Design of fabricated parabolic reflectors}

\begin{figure}
    \includegraphics[width=3.5in]{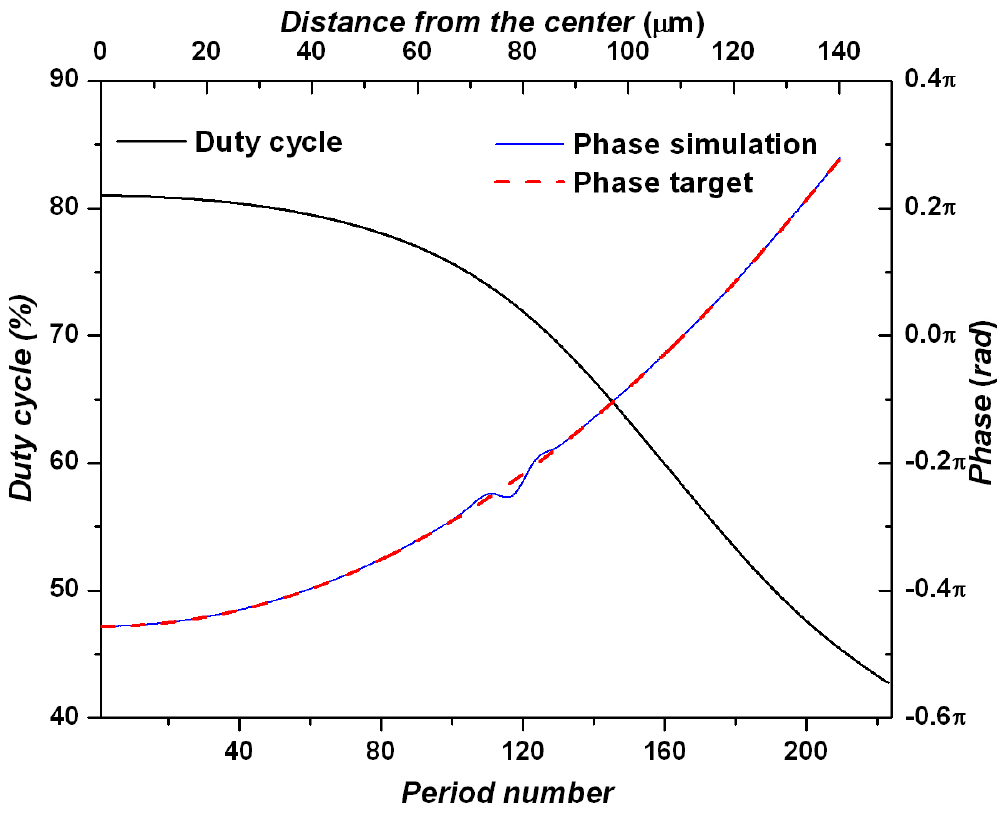}
    \caption{Solid black line: Local duty cycle (i.e., the ratio of groove width to period) as a function of distance from the center of the grating, designed for a focal length of 17.2~mm.  Blue solid line and red broken line: Expected (blue) and simulated (red) phase front of the reflected beam.  The incident field is a collimated beam having a waist of $100\mu$~m focused on the surface of the reflector.}
    \label{fig:wxy}
\end{figure}

Here we provide more detailed explanations of the designs of the parabolic reflectors shown in Figs.~3 and 4 of the main text.  It is basically an application of the discrete algorithm introduced above.  The design is based on a periodic grating of $450$~nm thick silicon with a $670$~nm period on a quartz substrate, which has the properties of the reflection coefficient shown in Fig.~\ref{fig:map_refl} (similar to that of Fig.~\ref{fig:phasecontour}, but for a different reference period).  We realized that varying the duty cycle while keeping the period fixed was sufficient for the parabolic focal lengths we wished to fabricate, corresponding to a design along a path parallel to the duty cycle axis (i.e., the broken white line in Fig.~\ref{fig:map_refl}, with the segment inside the black solid lines as the counterpart of Line 1 in Fig.~\ref{fig:phasecontour}).  For the cylindrical reflector, this resulted in uniform, parallel silicon grooves of different widths, with centers uniformed distributed along the $x$ direction with a period of 670~nm. A schematic of this pattern is shown in the middle of the first row of Fig.~4 in the main text, and the local duty cycle used in the design is shown in~Fig.~\ref{fig:wxy} as the solid black line.  For the spherical reflector, the result is a set of silicon grooves whose centers were also uniformly distributed along the $x$ direction with a period of $670\textrm{nm}$, but with each groove having a $y$-dependent width. This schematic is shown on the right of the first row of Fig.~4 in the main text.


We also used used the finite element method to perform a full-wave simulation of the designed cylindrical SWG mirror  ($f_x = 17.2$~mm, $f_y= +\infty$).  The targeted (parabolic) and simulated reflection phase profiles are shown in Fig.~\ref{fig:wxy} as the solid blue and broken red lines, respectively.  The expected and simulated phase profile agree very well, except for a small feature near period 120 where the grating reflectance of the period/duty cycle combination is slightly smaller (although still above 98\%) than the rest of the grating.


\bibliography{GMR_focus,qor_nps}


\end{document}